\documentclass[aps,prc,twocolumn,superscriptaddress,showpacs]{revtex4}

\usepackage{graphicx}
\usepackage{dcolumn}
\usepackage{bm}
\usepackage{ulem}
\usepackage{color}

\newcommand{\rag}{($\alpha$,$\gamma$)}

\newcommand{\rng}{(n,$\gamma$)}

\newcommand{\rpg}{(p,$\gamma$)}

\begin{document}

\title{
  Comment on ``Structure effects in the $^{15}$N(n,$\gamma$)$^{16}$N radiative
  capture reaction from the Coulomb dissociation of $^{16}$N''
}

\author{Peter Mohr}
\email{WidmaierMohr@t-online.de}
\affiliation{
Diakonie-Klinikum, D-74523 Schw\"abisch Hall, Germany}
\affiliation{
Institute for Nuclear Research (ATOMKI), H-4001 Debrecen, Hungary}

\date{\today}

\begin{abstract}
In their recent study 
Neelam, Shubhchintak, and Chatterjee
have claimed that ``it would
certainly be useful to perform a Coulomb dissociation experiment to find the
low energy capture cross section for the reaction''
$^{15}$N(n,$\gamma$)$^{16}$N. However, it is obvious that a Coulomb
dissociation experiment cannot constrain this capture cross section because
the dominating branchings of the capture reaction lead to excited states in
$^{16}$N which do not contribute in a Coulomb dissociation experiment. An
estimate of the total $^{15}$N(n,$\gamma$)$^{16}$N cross section from Coulomb
dissociation of $^{16}$N requires a precise knowledge of the $\gamma$-ray
branchings in the capture reaction. Surprisingly, the calculation of 
Neelam, Shubhchintak, and Chatterjee
predicts a strongly energy-dependent ground state branching of
the order of 0.05\,\% to 0.6\,\% at energies between 100 and 500\,keV which is
almost 2 orders of magnitude below calculations in the direct capture
model. Additionally, this calculation of 
Neelam, Shubhchintak, and Chatterjee
deviates
significantly from the expected energy dependence for $p$-wave capture.
\end{abstract}


\maketitle

Neelam, Shubhchintak, and Chatterjee (hereafter: NSC) \cite{Nee15} study the
Coulomb dissociation (CD) cross section of $^{16}$N and apply detailed balance
to derive the inverse $^{15}$N\rng $^{16}$N radiative capture cross
section. In detail, NSC calculate the CD for the four low-lying states in
$^{16}$N with $J^\pi = 2^-, 0^-, 3^-$, and $1^-$ which are located below
excitation energies of $E^\ast \approx 400$\,keV. NSC use spectroscopic
factors from the shell model taken from \cite{Mei96} which are close to unity
for the states under consideration. This is confirmed experimentally by
transfer data in \cite{Bar08} and by the analysis of neutron scattering
lengths \cite{Mohr97} but also lower values have been derived from transfer
\cite{Boh72,Guo14}. Spectroscopic amplitudes close to unity have also been
calculated recently in \cite{Pah15}.

Indeed, such a study can be made in theory; however, under normal experimental
conditions $^{16}$N is in its $2^-$ ground state, and thus a CD experiment is
only able to constrain the $^{15}$N(n,$\gamma_0$)$^{16}$N$_{\rm{g.s.}}$ cross
section but not the partial capture cross sections to the low-lying excited
states. Consequently, the conclusion of NSC ``to find the low energy capture
cross section'' from CD is misleading. Additionally, it will be difficult to
obtain a sufficient energy resolution in the CD experiment to derive the
low-energy capture cross section using Eq.~(3) of NSC \cite{Nee15}.

It is stated by NSC that CD theory has been used successfully to determine the
$^8$Li\rng $^{9}$Li \cite{Ban08} and $^{14}$C\rng $^{15}$C \cite{Shu14} cross
sections from CD of $^{9}$Li and $^{15}$C. Indeed, for these capture reactions
the ground state contributions are dominating, and thus experimental CD data
can be used to determine the total capture cross section. However, later NSC
claim that ``the Coulomb dissociation method has been used to find the neutron
capture cross section to different states of $^8$Li \cite{Izsak13} and also to
find the contributions of the projectile excited states in the charged
particle capture reactions \cite{Fleu05,Lang14}''. Ref.~\cite{Izsak13}
explicitly states that the experimental CD data for $^8$Li have to be
corrected for excited state contributions in the $^7$Li\rng $^8$Li reaction,
and this correction of about $10\,\% - 20$\,\% is estimated from experimental
branching ratios in the $^7$Li\rng $^8$Li capture reaction, see Eq.~(6) in
\cite{Izsak13}. Ref.~\cite{Fleu05} studies the $^{12}$C\rag $^{16}$O reaction
where the ground state contribution is dominating. Ref.~\cite{Lang14} analyzes
the $^{30}$S\rpg $^{31}$Cl reaction with its tiny $Q$-value of about
$+280$\,keV where the ground state of $^{31}$Cl is the only bound state. Thus,
none of Refs.~\cite{Izsak13,Fleu05,Lang14} matches the statement of NSC.

Experimental $^{16}$N CD data can indeed be used to determine the
$^{15}$N(n,$\gamma_0$)$^{16}$N$_{\rm{g.s.}}$ cross section. The determination
of the total $^{15}$N\rng $^{16}$N capture cross section is practically not
constrained by the small ground state contribution which is experimentally
accessible by CD. According to calculations (details see below), the
contribution of each low-lying excited state exceeds the ground state
contribution. Thus, the determination of the total capture cross section from
CD data requires additional information on the $\gamma$-ray branching ratios
in the $^{15}$N\rng $^{16}$N capture reaction.

\begin{figure}[htb]
\includegraphics[width=\columnwidth,clip=]{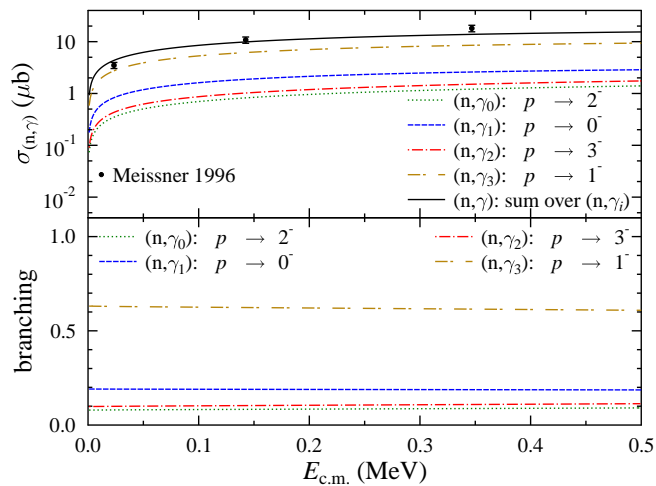}
\caption{
\label{fig:sigma}
(Color online)
Partial cross sections of $^{15}$N(n,$\gamma_i$)$^{16}$N (upper part) and
branching ratios $\sigma_{(\rm{n},\gamma_i)}/\sigma_{(\rm{n},\gamma)}$ (lower
part). The data are taken from \cite{Mei96}. Further 
discussion see text.
}
\end{figure}

Surprisingly, in the calculations of NSC this branching shows a significant
energy dependence (see Fig.~2 of \cite{Nee15}; note that the energies of the
experimental data points in the upper part differ from the lower part which is
probably the consequence of a missing conversion to the center-of-mass
system). This energy-dependent branching is a very unexpected result because
the energy dependence of the capture cross section at low energies is
typically governed by the angular momentum $l$ of the incoming neutron,
leading to an approximate $\sigma_{({\rm{n}},\gamma)} \sim E^{l-1/2}$
proportionality. For the transitions under consideration from incoming
$p$-waves to bound $s$- and $d$-states a similar energy dependence ($\sim
\sqrt{E}$) is expected, and the best description of the energy dependence in
direct capture (DC) calculations below about 500\,keV has been found
\cite{Mei96} with
\begin{equation}
\sigma_{({\rm{n}},\gamma)} = A \, E^{0.5} - B \, E^{1.2}
\label{eq:sigma}
\end{equation}
and the parameters $A$ and $B$ as given in Table III of \cite{Mei96}. Note
that $B < A$ which makes the second term in Eq.~(\ref{eq:sigma}) to a small
correction to the dominating $\sqrt{E}$ energy dependence for $E < 1$\,MeV.
These results are shown in Fig.~\ref{fig:sigma} together with the experimental
data of \cite{Mei96}. For easy comparison the same scale as in Fig.~2 of NSC
\cite{Nee15} has been chosen for the presentation of the cross sections.

The lower part of Fig.~\ref{fig:sigma} shows that the calculations of
\cite{Mei96} in the DC model lead to an almost energy-independent branching
ratio for the transitions from the incoming $p$-wave to the four bound $s$-
and $d$-states in $^{16}$N. The ground state contribution amounts to about
10\,\% in the whole energy range up to 500\,keV. Contrary to the DC
calculations of \cite{Mei96}, the new results of NSC show a significantly
steeper energy dependence for the transitions to the bound $d$-states ($2^-$,
$3^-$) with a ground state branching of about 0.05\,\% at 100\,keV and about
0.6\,\% at 500\,keV. This essential discrepancy of about 2 orders of
magnitude for the ground state branching ratio has to be well understood
before any conclusions on the total $^{15}$N\rng $^{16}$N capture cross
section can be drawn from experimental CD data of $^{16}$N and the model
calculations of NSC.

For completeness I point out that the expected weak energy dependence of the
branching ratios has also been found in DC calculations for the corresponding
transitions from incoming $p$-waves to bound $s$- and $d$-states in the
neighboring $^{14}$C\rng $^{15}$C \cite{Wie90} and $^{16}$O\rng $^{17}$O
\cite{Iga95,Men00} reactions. As Fig.~1 in \cite{Men00} shows, this weak
energy dependence of the branching ratio is also confirmed experimentally for
the $^{16}$O\rng $^{17}$O reaction.

In conclusion, contrary to the suggestion of NSC, the determination of the
total $^{15}$N\rng $^{16}$N capture cross section from a CD experiment on
$^{16}$N is not possible because the CD data are related only to the small
ground state branch in the capture reaction. The huge and surprising
discrepancy between the energy dependence of the branching ratios in the CD
calculations by NSC and in the DC calculations \cite{Mei96} (and similar
results in \cite{Wie90,Men00} for the neighboring $^{14}$C and $^{16}$O
nuclei) makes it impossible to correct experimental CD data of $^{16}$N for
the determination of the total $^{15}$N\rng $^{16}$N cross section using
theoretical branching ratios.

\smallskip
This work was supported by OTKA (K101328 and K108459).

\end{document}